\pdfoutput=1
\documentclass[11pt]{article}

\usepackage[preprint]{acl}
\usepackage{times}
\usepackage{latexsym}
\usepackage[T1]{fontenc}
\usepackage[utf8]{inputenc}
\usepackage{microtype}
\usepackage{inconsolata}
\usepackage{graphicx}
\usepackage{url}
\usepackage{subcaption}
\usepackage{pdflscape} %landscape orientation 
\usepackage{booktabs} % tables
\usepackage{multirow} % tables
\usepackage{adjustbox}
\usepackage{longtable}
\usepackage{xurl}

\title{ChunkNorris: A High-Performance and Low-Energy Approach to PDF Parsing and Chunking}

\author{Mathieu Ciancone \\
  \textit{Wikit} / Lyon, France \\
  \texttt{mathieu@wikit.ai} \\\And
  Clovis Varangot-Reille \\
  \textit{Wikit} / Lyon, France \\
  \textit{Laboratoire Hubert Curien }\\
  Université Jean Monnet \\ Saint-Etienne, France \\
  \texttt{clovis@wikit.ai} \\\And
  Marion Schaeffer \\
  \textit{Wikit} / Lyon, France \\
  \textit{INSA Rouen Normandie} \\ Rouen, France \\
  \texttt{marion.schaeffer} \\
  \texttt{@insa-rouen.fr}}

\begin{document}
\maketitle
\begin{abstract}
    In Retrieval-Augmented Generation applications, the Information Retrieval part is central as it provides the contextual information that enables a Large Language Model to generate an appropriate and truthful response. 
    High quality parsing and chunking are critical as efficient data segmentation directly impacts downstream tasks, i.e. Information Retrieval and answer generation. 
    In this paper, we introduce ChunkNorris, a novel heuristic-based technique designed to optimise the parsing and chunking of PDF documents. 
    Our approach does not rely on machine learning and employs a suite of simple yet effective heuristics to achieve high performance with minimal computational overhead. 
    We demonstrate the efficiency of ChunkNorris through a comprehensive benchmark against existing parsing and chunking methods, evaluating criteria such as execution time, energy consumption, and retrieval accuracy. 
    We propose an open-access dataset to produce our results. 
    ChunkNorris outperforms baseline and more advanced techniques, offering a practical and efficient alternative for Information Retrieval tasks. 
    Therefore, this research highlights the potential of heuristic-based methods for real-world, resource-constrained RAG use cases. The package is available at \href{https://github.com/wikit-ai/chunknorris}{https://github.com/wikit-ai/chunknorris}.
\end{abstract}

\section{Introduction}

    Retrieval-Augmented Generation (RAG) is an advanced paradigm in Natural Language Processing (NLP) that combines the strengths of Information Retrieval (IR) and generative models to address tasks requiring extensive knowledge and contextual understanding \citep{Lewis2020}. 
Unlike standalone generative models, RAG dynamically integrates external knowledge sources by retrieving relevant documents or data during inference\footnote{\url{https://arxiv.org/pdf/2312.10997}, accessed on \today.}. 
This retrieval step ensures the generated responses are coherent as well as grounded in up-to-date and accurate information, mitigating issues like hallucination \citep{Bouvard2024}.
Therefore, RAG is particularly valuable for applications such as question-answering, conversational agents, document summarisation, and decision support \citep{Fan2024}. 

The retrieval step in an RAG system queries a knowledge base, typically a large corpus of unstructured or semi-structured documents, to identify and extract the most relevant content for a given input \citep{Bouvard2024}. 
This process often involves techniques from the field of IR, such as lexical matching (e.g., TF-IDF \citep{Sparck1988}, BM25 \citep{Robertson1976}) or dense vector retrieval using embeddings generated by pre-trained language models (e.g., Sentence-BERT \citep{Reimers2019}). 
The retrieved results are ranked based on their relevance to the query and passed to the generative model as context \citep{Tao2023}.
Advanced systems may incorporate hybrid retrieval strategies, combining heuristic-based approaches with Machine Learning (ML) for improved performance. 

To feed the retriever of a RAG system, a knowledge base is constructed from documents that vary greatly in terms of format and complexity \citep{Zhang2024}.
To ensure good retrieval performance, documents must be parsed and chunked. 
Regarding PDF documents, parsing involves extracting structured and unstructured data from a format designed for human readability rather than machine processing, often requiring the handling of intricate layouts, multi-column text, tables, figures, and metadata\footnote{\label{Narayan2024} \url{https://arxiv.org/abs/2410.09871v1}, accessed on \today.}. 
The process typically begins with text extraction using libraries or tools such as \emph{PyPDF2}\footnote{\label{PyPDF2} \url{https://pypdf2.readthedocs.io/en/3.x/}, accessed on \today.}, \emph{PDFPlumber}\footnote{\url{https://github.com/jsvine/pdfplumber}, accessed on \today.}, or \emph{PyMuPDF}\footnote{\url{https://pymupdf.readthedocs.io/en/latest/}, accessed on \today.}, followed by additional preprocessing to clean and structure the extracted content. 
Key challenges include preserving semantic coherence, managing irregular formatting, and accurately reconstructing the document's logical flow\footref{Narayan2024}. 

Once the document structure and raw text are extracted from a document through parsing techniques, that content undergoes chunking which aims to segment the content into smaller, semantically coherent units, or \emph{chunks}, to facilitate efficient storage, indexing, and retrieval \cite{Kshirsagar2024}. 
This segmentation ensures that the retrieval system can identify and provide precise, contextually relevant information in response to a query rather than retrieving entire documents or unmanageable text blocks\footnote{\label{Yepes2024} \url{https://arxiv.org/html/2402.05131v3}}. 
Effective chunking strategies balance granularity, ensuring chunks are neither too large to dilute relevance nor too small to lose context\footref{Yepes2024}. 
Techniques for chunking often leverage heuristic rules, such as splitting by paragraph, sentence, or headings, while more advanced methods may incorporate semantic analysis to group related content meaningfully \cite{Kshirsagar2024}. 

Therefore, we aim to propose an efficient and low-energy solution for parsing and chunking PDF documents to improve IR performance: ChunkNorris. 
We benchmark ChunkNorris with existing parsing and chunking techniques on various criteria with an open-access dataset we build and propose to the community. 
The code for the benchmark is available in our GitHub repository:
\url{https://github.com/wikit-ai/bench-chunknorris-acl2025}

\section{Related Work}

    The PDF was invented to address the need to encapsulate documents to ensure readability across various platforms. 
Since its inception, continuous research has been conducted on how to effectively parse PDF files, given their complex structure. 
With the rise of Large Language Models (LLMs) and RAG applications, the need for efficient and accurate PDF parsing has become more critical. 
Significant advancements have been made in parsing techniques for PDF documents, driven by the increasing need to extract structured data from inherently unstructured or semi-structured content. 
State-of-the-art methods often combine traditional approaches with ML techniques to handle the complexities of the PDF format.
Traditional methods rely on libraries like \emph{PDFMiner}\footnote{\url{https://pdfminersix.readthedocs.io/en/latest/}, accessed on \today.} or \emph{PyPDF2}\footref{PyPDF2}, which provide programmatic access to text, images, and metadata. 
However, these tools only provide simple text extraction and do not provide information about document layout, hierarchical information, or structured information like tables. 
Heuristics-based methods were initially developed to enhance basic text extraction.
However, they tend to be replaced in favour of computer vision approaches that leverage ML to improve performance. 

Models trained with annotated data integrate textual and layout information, enabling improved tables, forms, and complex structures extraction. 
This is the case for Python librairies sush as \emph{Unstructured}\footnote{\url{https://docs.unstructured.io/welcome}, accessed on \today.} and \emph{Docling}\footnote{\url{https://ds4sd.github.io/docling/}, accessed on \today.}. 
Furthermore, hybrid techniques combine multiple approaches, such as \emph{Open-Parse}\footnote{\url{https://github.com/Filimoa/open-parse}, accessed on \today.}, which uses mainly heuristics, and ML as an option; \emph{Marker}\footnote{\url{https://github.com/VikParuchuri/marker}, accessed on \today.}, which uses ML, Optical Character Recognition (OCR), and has LLM support; and \emph{LLM Sherpa}\footnote{\url{https://github.com/nlmatics/llmsherpa}, accessed on \today.}, which uses heuristics and LLM techniques. 
While such techniques provide high-quality results, their use in production is limited due to high processing time, computational resources and annotation requirements that often do not match the constraints of production environments.

Regarding chunking, the literature suggests two possible scenarios. 
The first is where the parsing and chunking methods are separate. 
For example, \emph{PyPDF2}, \emph{Unstructured}, \emph{Marker}, \emph{LLM Sherpa} and \emph{NV-Ingest}\footnote{\url{https://github.com/NVIDIA/nv-ingest}, accessed on \today.} are parsing-only tools. 
Their output can be processed using independent chunking techniques, such as the \emph{LangChain's} commonly used \emph{recursive character text splitter}\footnote{\url{https://python.langchain.com/docs/how\_to/recursive\_text\_splitter/}, accessed on \today.}. 
In the second case, the parsing and chunking methods are built together as a pipeline to combine these two steps effectively. 
This is the case, for example, for \emph{Open-Parse} and \emph{Docling}. 
Among the most common chunking methods is length-based splitting, which segments text based on specified size limits, such as tokens or characters. 
Token-based splitting is particularly useful when interfacing with language models, as it aligns with their input constraints, while character-based splitting ensures consistency across diverse text types \cite{Kshirsagar2024}. 
These methods are straightforward to implement, adaptable, and produce uniform chunk sizes. 
A more advanced approach is hierarchical splitting, which leverages the natural structure of text, such as paragraphs, sentences, and words, to create coherent splits \cite{Kshirsagar2024}.
Tools like LangChain's recursive character text splitter\footnote{\url{https://python.langchain.com/docs/how_to/recursive_text_splitter/}, accessed on \today.} exemplify this technique by prioritising larger units (e.g., paragraphs) and recursively splitting smaller units when necessary, preserving the semantic flow of the text. 
For documents with inherent structures, such as HTML, Markdown, or JSON, structure-based splitting leverages these formats' structural information, such as headers, tags, or object boundaries, to create contextually rich chunks \cite{Kshirsagar2024}. 
This approach maintains the document's logical organisation and is particularly effective for preserving semantic relationships. 
Finally, semantic-based splitting goes a step further by analysing the content's meaning to identify significant shifts in context \cite{Kshirsagar2024}. 
This last method often uses sliding window techniques and embeddings to detect breakpoints in the text, ensuring chunks remain semantically coherent. 
While semantic-based splitting stands out for its ability to directly analyse and maintain contextual integrity, it requires much more execution time and computational resources that are not always available in practice.

With ChunkNorris, we propose an efficient and unsupervised ML-free parsing and chunking method. 
We aim to ensure fast document ingestion with limited computational resources while getting the most out of the document structure.
Our algorithm is robust and enables coherent chunking of various documents.

\section{ChunkNorris}

    This work introduces ChunkNorris, a novel parsing and chunking algorithm designed to efficiently process documents without requiring GPU acceleration. 
While chunking based on document titles has demonstrated high effectiveness, it has not been widely applied to PDFs, primarily because detecting headers and their hierarchy from document layout is challenging.
We developed ChunkNorris, which leverages title-based segmentation to produce high-quality chunks while remaining lightweight and efficient.

\subsection{Parser}

    A parser is a computational tool or algorithm that analyses and processes structured or unstructured data, converting it into a machine-readable format. 
    Parsers are essential components of text processing systems that clean and format input documents. 
    ChunkNorris parser's primary role is to take a file or a string as input and produce a clean, markdown-formatted output suitable for further processing by a chunker. 
    Currently, ChunkNorris supports three parsers: MarkdownParser, HTMLParser, and PdfParser. 
    Regardless of the input type, all parsers generate a unified MarkdownDoc object, which serves as input for a chunker.
    This work focuses on the PdfParser, which is designed to extract and structure content from PDF files. 

    ChunkNorris implementation relies on the \emph{PyMuPDF} library. This tool provides fast implementations of a great variety of utility functions for document processing.
    The parsing process begins by opening the PDF using \emph{PyMuPDF}. Text is retrieved in the form of spans, which are defined as sets of consecutive characters sharing the same formatting properties. 
    Based on their attributes and location, spans undergo a series of processing steps to ensure accurate structuring of the document.\\
    \textbf{Headers and footers:} If a span's bounding box appears in the same position on more than 33\% of the document's pages, it is flagged as a header or footer and subsequently removed.\\
    \textbf{Links:} In PDF documents, hyperlinks exist as invisible clickable boxes layered over spans. 
    The PdfParser retrieves links and binds them to their corresponding spans to avoid loosing that information. To our knowledge, none of the other existing PDF parsing tools extract hyperlinks.\\
    \textbf{Tables:} Tables in PDFs vary widely in layout, making them challenging to parse. 
    They fall into three categories: those with visible cell boundaries, those inferred from content alignment, and those embed as images. 
    In the first case, the table structure is represented by line vectors, which can be recombined for parsing. 
    ChunkNorris employs a vectorised line recombination method for efficient table structure extraction. 
    It also handles the parsing of tables with merged cells.\\
    \textbf{Lines/Blocks:} Following text extraction, spans are grouped to form coherent text units. 
    Consecutive spans on the same vertical position are merged into lines. 
    Lines are further grouped into blocks, which may represent a paragraph or a section title. 
    Building of blocks is based on the line spacing of the document's body content, which refers to the vertical distance between bounding boxes of consecutive lines.\\
    \textbf{Main title:} The parser also attempts to infer the document's main title by analysing blocks on the first page. 
    Blocks with font sizes larger than the body text are considered potential title candidates.\\
    \textbf{Section headers:} The PdfParser detects section headers and their hierarchy. 
    It first checks the document metadata for a Table of Contents (ToC). 
    If found, it is used directly.
    Otherwise, the parser searches for a structured ToC within the document using regular expressions. 
    Header levels are inferred from indentation (deeper levels are further right) or numbering patterns (e.g., \emph{1.}, \emph{1.1}, \emph{1.1.a}). 
    If no ToC is available, font sizes determine hierarchy, with smaller fonts indicating deeper levels.\\
    \textbf{Other:} Various document attributes are also extracted to characterise the content, such as document orientation, font size of the body, etc.\\
    Finally, after completing the extraction and structuring process, the PdfParser generates a markdown-formatted document, ensuring a clean and structured representation of the original PDF content ready to be processed by a chunker or other text-processing modules.
    
\subsection{Chunker}

    Chunkers process parser outputs by segmenting them into coherent units.
    In ChunkNorris, all parsers produce Markdown-formatted MarkdownDoc objects, ensuring compatibility with the MarkdownChunker.
    Markdown is the chosen standard for its readability by both humans and LLMs while offering sufficient structure for chunking.
    As a result, the MarkdownChunker is currently the only implemented chunker, though others may be developed if new parser output formats emerge.

    The chunking strategy employed by the MarkdownChunker is based on several guiding principles. 
    First, each chunk must contain homogeneous information.
    Therefore, the chunking process relies on document section headers to define chunk boundaries. 
    Second, each chunk must retain contextual information, as sections of a document can lose meaning when read in isolation.
    To preserve context, the headers of all parent sections are prepended to each chunk.
    Third, chunk sizes should be as uniform as possible.
    Embedding models used in IR are sensitive to chunk length, resulting in higher embedding similarity for chunks of similar length to the query. 
    If a chunk is significantly longer than the query, its similarity score may decrease despite relevance. 
    Chunkers aim to maintain a consistent chunk size whenever possible to mitigate this issue.
    
    The chunking process begins by constructing a ToC tree from document headers.
    Chunks are then recursively generated based on the ToC structure, with each chunk containing the titles of upper sections and the content of the current section.
    A chunk is subdivided using available subsections if it exceeds the soft word limit.
    Otherwise, it remains intact.
    After chunking, refinements are applied: sections below a minimum word count are discarded to ensure only meaningful chunks are retained.
    Chunks exceeding a hard limit are split into subchunks at newline characters, which ensure tables and code blocks remain within a single chunk.
    Titles from the original chunks are preserved at the beginning of each subchunk to maintain context.
    
    The final output of the MarkdownChunker is a list of Chunk objects, each containing its processed text, the associated parent headers, the starting line of the chunk within the original markdown-formatted document and, when relevant, the start and end page numbers of the paginated source file.
    This structured output ensures very fast and efficient document segmentation while preserving readability and contextual integrity.
    ChunkNorris parsers and chunkers can be wrapped up into pre-built pipelines. 
    They allow processing of documents with minimum code while ensuring constant output quality.

\section{Benchmark of parsing and chunking techniques}

    To evaluate ChunkNorris, we propose comparing its performance to other popular tools. 
This section presents the methodology we apply to compare ChunkNorris and the dataset constructed for the evaluation. 
The dataset is available as open-source on Hugging Face:
\url{https://huggingface.co/datasets/Wikit/PIRE}.

\subsection{Dataset}
\label{sec:dataset}

    We evaluate parsing and chunking in the context of RAG.
    Therefore, we construct the PDF dataset for Information Retrieval Evaluation (PIRE) designed explicitly for the IR use case to assess the parsing and chunking methods. 
    This dataset comprises 100 PDFs, combining 50 documents from the existing DocLayNet dataset \cite{Pfitzmann2022} and 50 newly collected PDFs whose diversity reflects real-world use cases. 
    The newly collected set includes 5 arXiv papers, 2 financial reports, 4 infographics, 4 legal documents, 3 IT documentations, 4 news articles, 3 PowerPoint-like documents, 13 PubMed papers, 3 organisation reports, 5 user manuals, and 5 Wikipedia articles.
    The documents were selected with consideration of their licenses. A list of all the PDFs and their references can be found in the Appendix \ref{sec:dataset-description}.
    This diverse selection ensures a broad range of document structures and content types, making it well-suited for evaluating the robustness of our approach.\\
    \subsubsection*{Single-chunk dataset}
    We first annotate three questions per PDF, resulting in 300 question-document pairs.
    Three annotators followed a structured methodology to identify the minimal passage within each document that contained the answer.
    A retrieval step was carried out to ensure only one passage contained the answer to each question.
    In this step, documents were parsed using the ChunkNorris parser and segmented into fixed-size chunks of 250 words.
    We then embed all the chunks using \emph{Alibaba-NLP/gte-large-en-v1.5}\footnote{\url{https://huggingface.co/Alibaba-NLP/gte-large-en-v1.5}, accessed on \today.}, a high-performing embedding model from the MTEB leaderboard \cite{Muennighoff2023} that remains computationally efficient with fewer than 500M parameters. 
    We compute the cosine similarity between the annotated question and all chunks, retrieving the top 10 highest-scoring chunks. 
    The annotators validate the question if no additional relevant chunks are found among these top results. 
    Otherwise, they refine or rewrite the question to better isolate a single relevant passage.
    We refer to this subset as \emph{single-chunk dataset}.\\
    \subsubsection*{Multi-chunk dataset}
    Additionally, we extend the dataset with another 32 questions, which require several pieces of information spread over multiple pages to be answered.
    This time, each question is matched with the pertinent passages of the corpus along with their source document and page.
    This part of the dataset is deliberately more complex than the first.
    We refer to it as \emph{multi-chunk dataset}.

    \paragraph{}
    This dataset provides a high-quality benchmark for evaluating parsing and chunking strategies in an IR context. 
    Incorporating diverse document types, structured annotations, and a robust validation process allows for a comprehensive assessment of our approach's in real-world retrieval scenarios.

\subsection{Evaluation methodology}

    To evaluate ChunkNorris, we compare its performance with other tools in the literature. 
    We choose methods for parsing and/or chunking PDF documents to carry out our benchmark. 
    After a wide-range screening of existing methods, we select those that show remarkable performance and have aroused great interest in the community.   
    For the parsing step, we compare \emph{Marker}, \emph{Open-Parse} and \emph{Docling}. \emph{PyPDF} is used as a baseline, as it can only perform text extraction without properly parsing the document. 
    All parsers are run with their default configuration. OCR is deactivated to avoid influencing the results, as it is not needed for the studied documents.
    We associate all these tools with two different chunking strategies: by page and with the recursive text splitter set to a size limit of 4000 characters per chunk and an overlap of 200 characters between chunks.
    Custom chunking strategies are available for \emph{Open-Parse} and \emph{Docling}, so we add them to the comparison.
    Additionally, \emph{Open-Parse} proposes two parsing backends: one using \emph{PyMuPDF}'s built-in functions and another leveraging Unitable \cite{unitable}, an ML framework for table extraction.
    The first one runs on CPU only while the latter demands a GPU. In the rest of this paper, they will be referred to as \emph{Open-Parse-P} and \emph{Open-Parse-U}, respectively.
    Appendix \ref{sec:parsers-features} summarises the features of the compared parsers. 

    We evaluate the pipelines based on various criteria, including execution time and environmental impact measured through energy consumption.
    For the latter, we use \emph{CodeCarbon}\footnote{\url{https://codecarbon.io/}, accessed on \today.} to measure directly electricity used by the GPU and \emph{psutil}\footnote{\url{https://psutil.readthedocs.io/en/latest/}, accessed on \today} to get the load percentage of CPU.
    CPU energy consumption is then calculated as :
    \begin{equation}
        E = CPU \; load \times time \times P
    \end{equation}
    where E is the energy consumed (Wh), computed by multiplying the CPU load percentage with the execution time (h) for parsing the dataset, and P (W) the CPU power provided by the manufacturer. 

    We use a retrieval task to evaluate the different parsing and chunking pipelines.
    We compare various embedding models listed in Table \ref{tab:model-characs} to avoid bias toward a specific chunking strategy.
\begin{table}
  \centering
  \begin{tabular}{llc}
    \hline
    \textbf{Provider} & \textbf{Model name} & \textbf{\#Params}\\
    \hline
    Snowflake & arctic-embed-xs & 23M \\
    & arctic-embed-m-v1.5 & 109M\\
    & arctic-embed-m-v2.0 & 305M\\
    BAAI & bge-small-en-v1.5 & 33M \\
    & bge-base-en-v1.5 & 109M\\
    & bge-large-en-v1.5 & 335M\\
    \hline
  \end{tabular}
  \caption{
    Embedding models used to assess retrieval performance. All models are available on HuggingFace.}
  \label{tab:model-characs}
\end{table}
    After parsing and chunking, we embed the chunks with all models.
    We are interested in the trade-off between the complexity of the parsing techniques and the size of the embedding models required to maximise IR task performance while minimising ecological impact.
    We use annotated questions of the dataset described in Section \ref{sec:dataset} to perform chunk retrieval using the cosine similarity between the question embedding and the chunk embeddings.
    We compute the recall and the Normalized Discounted Cumulative Gain (NDCG) for the 10 high-ranking chunks (@10).

\section{Results}

    \subsubsection*{Which parser is best suited to a production environment?}

\begin{table}[h!]
  \centering
  \begin{tabular}{lc}
    \hline
    \textbf{Parsers} & \textbf{Parsing time} \\
    \hline
    ChunkNorris & \underline{105 ms $\pm$ 296}\\
    Docling & 533 ms $\pm$ 1284 \\
    Marker & 717 ms $\pm$ 594  \\
    Open-Parse-P & 459 ms $\pm$ 485 \\
    Open-Parse-U & 2538 ms $\pm$ 2659 \\
    PyPDF & \textbf{91 ms $\pm$ 169} \\
    \hline
  \end{tabular}
  \caption{
    Average of the page parsing time for each parser. Hardware: CPU \emph{Intel(R) Xeon(R) Gold}; GPU \emph{Tesla V100S}; RAM 40 GiB.
    }
  \label{tab:parsers-time-per-page}
\end{table}

\begin{table}[h!]
  \centering
  \begin{tabular}{lcc}
    \hline
    \textbf{Parsers} & \begin{tabular}{@{}c@{}}\textbf{CPU}\\ \textbf{energy}\end{tabular} & \begin{tabular}{@{}c@{}}\textbf{GPU}\\ \textbf{energy}\end{tabular}\\
    \hline
    ChunkNorris & \underline{0.47 Wh} & \textbf{0.0 Wh}\\
    Docling & 5.11 Wh & \underline{23.24 Wh} \\
    Marker & 1.92 Wh & 58.09 Wh \\
    Open-Parse-P & 1.87 Wh & \textbf{0.0 Wh}\\
    Open-Parse-U & 34.32 Wh & 777.93 Wh \\
    PyPDF & \textbf{0.28 Wh} & \textbf{0.0 Wh}\\
    \hline
  \end{tabular}
  \caption{
    Energy consumption required to parse the 100 PDFs (5286 pages). Hardware: CPU \emph{Intel(R) Xeon(R) Gold}; GPU \emph{Tesla V100S}; RAM 40 GiB.}
  \label{tab:parsers-conso-total}
\end{table}

We first evaluate the average parsing time of a PDF page for each parsers in Table \ref{tab:parsers-time-per-page}.
Additionally, we compare CPU and GPU energy needed to run each parser on the entire dataset (100 PDFs and 5286 pages) in Table \ref{tab:parsers-conso-total}. 
We do not measure the execution time and energy consumed in the chunking stage, as it is negligible compared with the parsing stage. 
The experiment has been reproduced on another hardware, and results are displayed in Appendix \ref{tab:parsers-conso-total-local-env} to confirm the ranking.
As expected, the baseline \emph{PyPDF} is the fastest parser (91~ms per page) and consumes the least resources, with only 0.28~Wh to parse the entire dataset.
ChunkNorris is second for both criteria. With a parsing time close to 100~ms per page, it is an interesting asset for production environments, which are often subject to high ingestion workloads.
Regarding parsing time, \emph{Open-Parse-P}, \emph{Docling} and \emph{Marker} respectively rank next, with values around 500~ms per page.
However, they display significant differences in energy consumption. \emph{Open-Parse-P} remains within the low range of energy consumption with 1.92~Wh. In contrast, \emph{Docling} and \emph{Marker} show much higher values of around 28~Wh and 60~Wh, respectively, due to their requirement for GPU acceleration.
Finally, \emph{Open-Parse-U} performs significantly worse, showing much higher execution time and energy consumption than the other GPU-accelerated parsers.
This technique is hardly suitable for production environments, as it takes an average of more than 2.5~s to parse a PDF page.

\begin{figure*}[htbp]
  \begin{subfigure}{.48\linewidth}
    \centering\includegraphics[width=.98\linewidth]{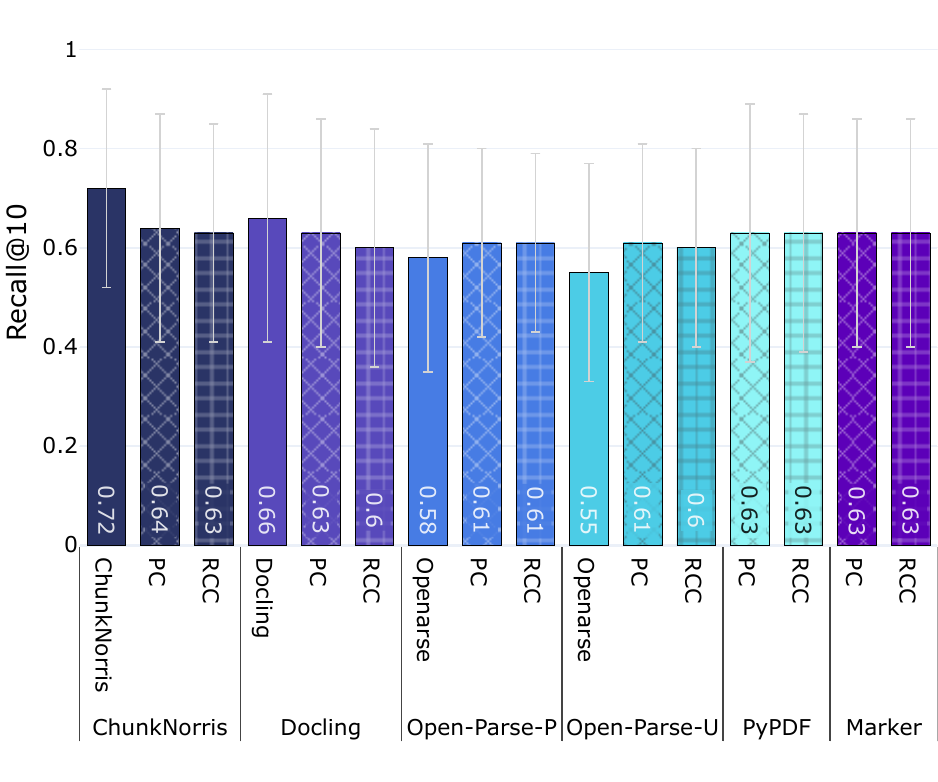}
    \caption{Single-chunk dataset}
  \end{subfigure}
  \begin{subfigure}{.48\linewidth}
    \centering\includegraphics[width=.98\linewidth]{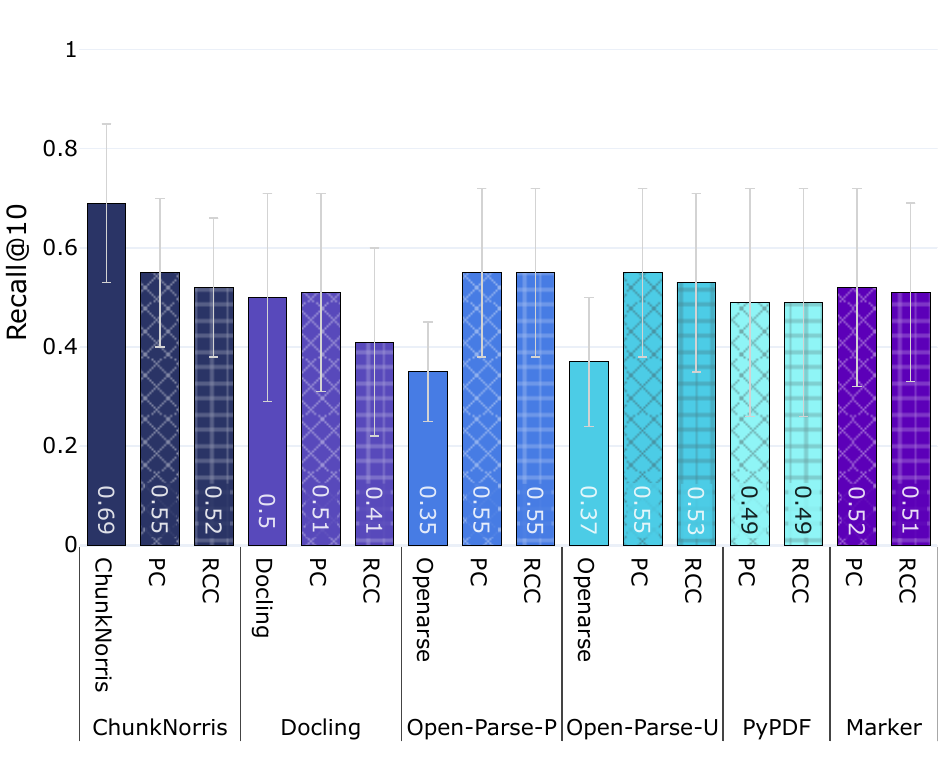}
    \caption{Multi-chunk dataset}
  \end{subfigure}
  \caption {Average recall@10 over all embeddings models for parsing and chunking pipelines. PC stands for Page Chunker, and RCC for Recursive Character Chunker.}
  \label{fig:recall-ir}
\end{figure*}

\begin{figure*}[h!]
  \begin{subfigure}{.48\linewidth}
    \centering\includegraphics[width=.98\linewidth]{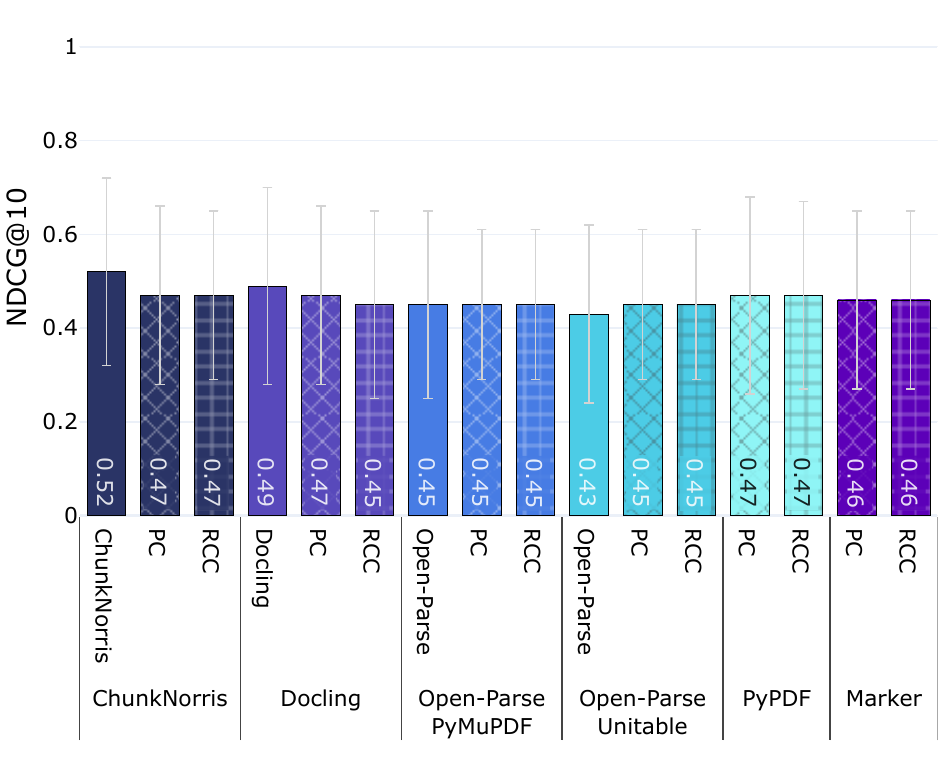}
    \caption{Single-chunk dataset}
  \end{subfigure}
  \begin{subfigure}{.48\linewidth}
    \centering\includegraphics[width=.98\linewidth]{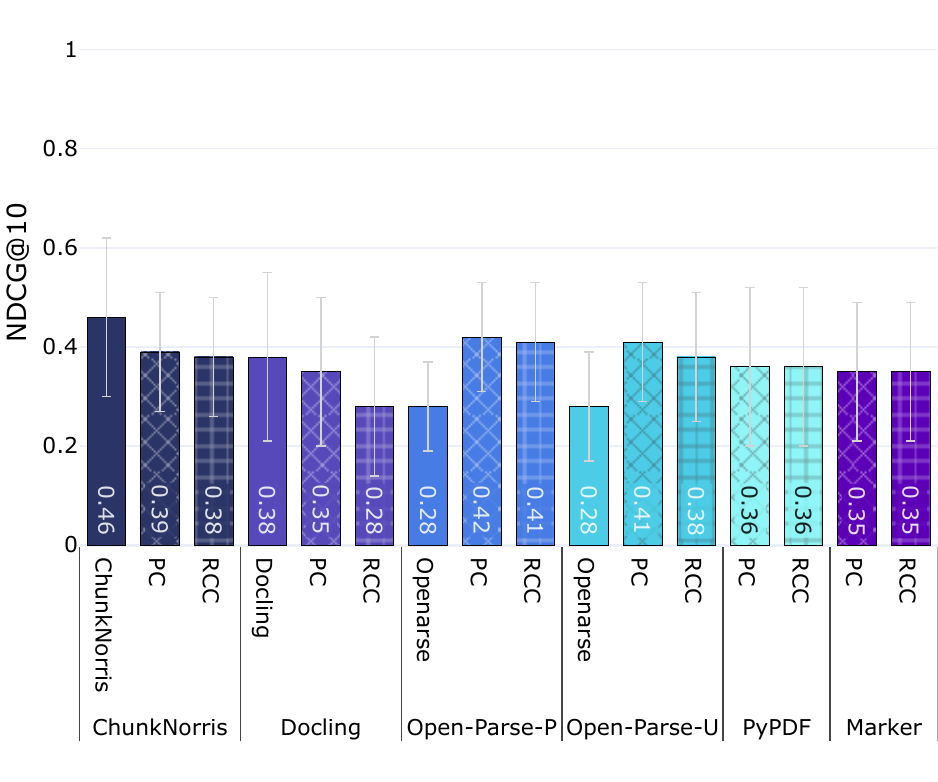}
    \caption{Multi-chunk dataset}
  \end{subfigure}
  \caption {Average NDCG@10 over all embeddings models for parsing and chunking pipelines. PC stands for Page Chunker, and RCC for Recursive Character Chunker.}
  \label{fig:ndcg-ir}
\end{figure*}

\subsubsection*{Which pipeline performs best for the IR task?}
Next, we evaluate each combination of parser and chunker on the IR task.
Figure \ref{fig:recall-ir} presents the average recall@10 across all embedding models.
The ChunkNorris pipeline performs best on single-chunk and multi-chunk datasets, highlighting its overall effectiveness.
However, apart from ChunkNorris, the methods' ranking differs between the two datasets.
\emph{Docling} is the second-best performer for the single-chunk dataset, followed by \emph{PyPDF} and \emph{Marker}. 
In contrast, for the multi-chunk dataset, \emph{Open-Parse} ranks second, followed by \emph{Marker} and \emph{Docling}. 
These results suggest that \emph{PyPDF} is well-suited for handling simple retrieval but struggles with more complex ones where information is scattered across multiple pages. In the meantime, \emph{Open-Parse} with page chunking is more effective when dealing with multi-chunk document retrieval.
Notably, ChunkNorris maintains strong and consistent performance across both scenarios, demonstrating its robustness regardless of retrieval difficulty.
In Figure \ref{fig:ndcg-ir}, we present the results for NDCG@10, which evaluates the ranking quality of the retrieved chunks. 
These results correlate with the recall and further confirm the strong performance of the ChunkNorris pipeline, which achieves the highest scores for both single-chunk and multi-chunk datasets.
The differences between pipelines are relatively small for the single-chunk dataset, indicating that most methods rank retrieved chunks similarly when dealing with simpler document structures. 
However, the contrast between results is more pronounced in the multi-chunk dataset.
\emph{Open-Parse} seems to stand out from other tools with page and recursive character chunkers, making it more suitable for complex multi-chunk retrieval.

We evaluate the performance variation across various parser and chunker combinations to analyse the impact of parsing and chunking separately based on the recall in Figure \ref{fig:recall-ir}. 
We observe that the performance of a single parser varies more significantly depending on the chunker used than the performance of different parsers with the same chunker. 
It suggests that the interaction between parser and chunker plays a crucial role rather than one component being universally more important than the other.
A clear example is the ChunkNorris parser, which no longer stands out from other pipelines when used without its dedicated chunker. 
This highlights that its strong performance stems from the synergy between its parser and chunker rather than in document parsing alone.
The exception to this trend is observed in the \emph{Open-Parse} pipeline on the multi-chunk dataset, where it performs significantly worse than other chunkers.

\subsubsection*{Which pipeline is the most robust to different embedding models?}
We now consider the results for each embedding model detailed in Appendix \ref{sec:pipeline-results}.
We compare Arctic\footnote{\url{https://huggingface.co/collections/Snowflake/arctic-embed-661fd57d50fab5fc314e4c18}, accessed on \today.} and BGE \cite{Xiao2024} models across three model sizes.
We first focus on the single-chunk dataset, and the results are presented in Table \ref{tab:comparison-single}.
The ChunkNorris pipeline consistently achieves the best recall across all model sizes for the Snowflake models, while NDCG remains close between ChunkNorris and \emph{Open-Parse}. 
As expected, the largest model yields the best retrieval results. 
An unexpected trend emerges: the middle-sized model performs worse than the small model, suggesting insufficient training or adaptation for the retrieval task.
For the BGE models, ChunkNorris performs best for the small and large models, while the middle-sized model favours \emph{Docling}. 
This indicates that while ChunkNorris remains robust across different embedding model sizes, some variations in performance emerge based on specific model architectures.
We then focus on the multi-chunk dataset with results presented in Table \ref{tab:comparison-multi}.
ChunkNorris dominates across all BGE models. 
\emph{Marker} performs particularly well in recall, while \emph{Open-Parse} shows strong results in NDCG, indicating that the resulting chunks allow for effective ranking. 
The performance consistency across all BGE model sizes is remarkable, making the smallest model an attractive choice for future use due to its efficiency in resource consumption without compromising performance.
ChunkNorris achieves the best recall with the smaller Snowflake model, excels in both recall and NDCG with the middle model, and leads in NDCG with the large model. 
When not leading, it closely competes with \emph{Open-Parse}. 
However, the middle Snowflake model again underperforms, mirroring its results on the single-chunk dataset. 
This anomaly suggests that model size alone does not dictate performance, and specific training dynamics may influence retrieval effectiveness.

\section{Conclusion}

    We propose ChunkNorris, a fast and reliable parsing and chunking tool for PDF documents. 
We demonstrate ChunkNorris' interest over other popular tools across production constraints such as execution time, resource consumption, and IR performance.
We ensure a robust and comprehensive comparison by testing on a diverse set of documents and embedding models.
ChunkNorris demonstrates outstanding performance, consistently surpassing other methods in recall and ranking quality while maintaining lower parsing time and energy consumption. 
Remarkably, it achieves this efficiency without significant trade-offs, staying close to the computational requirements of simple plain text extraction.
These results highlight ChunkNorris efficiency, making it particularly well-suited for ingestion pipelines handling heavy workloads.
Ongoing work focuses on enhancing the parsing of specific PDF components, such as advanced table layouts or mixed-up reading orders, while maintaining speed and reliability. 
By releasing ChunkNorris as an open-source Python package, we aim to simplify PDF parsing and chunking while reducing its ecological impact.
The benchmark proposed is a first attempt to compare parsing and chunking tools. The code is designed for easy extensibility, allowing to include additional methods such as \emph{Unstructured}\footnote{\url{https://github.com/Unstructured-IO/unstructured}, accessed on \today.} and \emph{NV-Ingest}\footnote{\url{https://github.com/NVIDIA/nv-ingest}, accessed on \today.}.

\section*{Limitations}

Our study has some limitations that should be considered to expend this work.
First, we use default settings for all pipelines because we leave it up to each tool to define the most relevant configurations. In actual use cases, users of these tools rarely try to optimise the various parameters. However, further optimisation could improve performance for some methods.
Then, the dataset we propose is limited by its size, especially the subset of questions requiring multiple chunks for retrieval. 
A more extensive and diverse dataset would strengthen the impact and generalisability of our findings.
Another limitation concerns our evaluation, which focuses solely on the RAG use case. 
Exploring other use cases to evaluate parsing and chunking techniques would be interesting. 
More specifically, assessing the quality of parsing, for example, with structured data extraction or document layout analysis, would be interesting. 
Finally, ChunkNorris follows a right-to-left, top-to-bottom reading order, which may limit its effectiveness for multilingual applications, particularly for languages with different text orientations or complex layouts. 
Future work should explore methods to adapt or extend ChunkNorris for broader language support. 

\bibliography{bibliography}

\appendix

\onecolumn
\section{Dataset description}
\label{sec:dataset-description}
\begin{longtable}{cp{8cm}c}
    \hline
    \textbf{Category} & \textbf{URL} & \textbf{Consultation date} \\
    \hline
        Arxiv paper & \url{https://arxiv.org/pdf/2501.01818} & 09/01/2025\\
    Arxiv paper & \url{https://arxiv.org/pdf/2002.12327} & 09/01/2025\\
    Arxiv paper & \url{https://arxiv.org/pdf/2404.08471} & 09/01/2025\\
    Arxiv paper & \url{https://arxiv.org/pdf/2409.14160} & 09/01/2025\\
    Arxiv paper & \url{https://dl.acm.org/doi/pdf/10.1145/3442188.3445922} & 09/01/2025\\
    Financial report & \url{https://upload.wikimedia.org/wikipedia/foundation/f/f6/Wikimedia_Foundation_2024_Audited_Financial_Statements.pdf} & 10/01/2025\\
    Financial report & \url{https://www.sec.gov/Archives/edgar/data/1318605/000162828024043486/tsla-20240930.htm} & 10/01/2025\\
    Infographic & \url{https://www.ccpl47.fr/wp-content/uploads/2024/02/BD-EN_calendrier-Lauzun-2024.pdf} & 10/01/25\\
    Infographic & \url{https://commons.wikimedia.org/wiki/File:ORCID_Infographic_2019.pdf} & 09/01/25\\
    Infographic & \url{https://github.com/wikit-ai/olaf/blob/main/docs/Poster_OLAF_2023.pdf} & 09/01/25 \\
    Infographic & \url{https://upload.wikimedia.org/wikipedia/commons/9/9e/Understanding_Creative_Commons_license_%28infographic%29.pdf} & 10/01/25\\
    Legal document & \url{https://datamillnorth.org/download/vdwno/e36a9342-4b29-4638-86a9-572acb66469d/ukcp18-project-technical-overview_July.pdf} & 09/01/25 \\
    Legal document & \url{https://assets.publishing.service.gov.uk/government/uploads/system/uploads/attachment_data/file/493524/horr90-opiate-crack-cocaine-users.pdf} & 09/01/25 \\
    Legal document & \url{https://assets.publishing.service.gov.uk/government/uploads/system/uploads/attachment_data/file/380586/prison-population-projections-2014-2020.pdf} & 09/01/25\\
    Legal document & \url{https://eur-lex.europa.eu/resource.html?uri=cellar:a3c806a6-9ab3-11ea-9d2d-01aa75ed71a1.0001.02/DOC_1&format=PDF} & 09/01/25\\
    Microsoft documentation & \url{https://learn.microsoft.com/pdf?url=https%3A%2F%2Flearn.microsoft.com%2Fen-us%2Foffice%2Fpdf%2Ftoc.json} & 10/01/25\\
    Microsoft documentation & \url{https://docs.aws.amazon.com/pdfs/serverless/latest/devguide/serverless-core.pdf#welcome} & 10/01/25\\
    Microsoft documentation & \url{https://fr.slideshare.net/slideshow/welcome-to-word-template/250355217} & 10/01/25\\
    News & \url{https://about.newsusa.com/new-artificial-intelligence-summit-series-begins-with-energy} & 14/01/25\\
    News & \url{https://www.newscanada.com/en/three-ways-canadian-communities-are-reducing-flood-risks-139844} & 14/01/25 \\
    News & \url{https://about.newsusa.com/3-great-resources-to-kick-start-your-financial-planning-career} & 14/01/25\\
    News & \url{https://www.newscanada.com/en/the-top-ai-powered-tech-trends-in-2025-139854} & 14/01/25 \\
    Powerpoint like & \url{https://query.prod.cms.rt.microsoft.com/cms/api/am/binary/RE4X6Ux} & 10/01/25\\
    Powerpoint like & \url{https://wiki.creativecommons.org/images/8/88/Publicdomain.pdf} & 13/01/25\\
    Powerpoint like & \url{https://download.microsoft.com/download/1/2/6/1269C192-F79E-4918-B737-D828E0522349/Word%20QS.pdf} & 10/01/25 \\
    PubMed paper & \url{https://pmc.ncbi.nlm.nih.gov/articles/PMC10681710/pdf/niad025.pdf} & 09/01/25\\
    PubMed paper & \url{https://pmc.ncbi.nlm.nih.gov/articles/PMC11562755/pdf/jop-165-2863.pdf} & 09/01/25 \\
    PubMed paper & \url{https://pmc.ncbi.nlm.nih.gov/articles/PMC11693606/pdf/41586_2024_Article_8275.pdf} & 09/01/25\\
    PubMed paper & \url{https://pmc.ncbi.nlm.nih.gov/articles/PMC11537970/pdf/41593_2024_Article_1741.pdf} & 09/01/25\\
    PubMed paper & \url{https://pmc.ncbi.nlm.nih.gov/articles/PMC6102917/pdf/12913_2018_Article_3470.pdf} & 09/01/25 \\
    PubMed paper & \url{https://pmc.ncbi.nlm.nih.gov/articles/PMC11638540/pdf/main.pdf} & 13/01/25\\
    PubMed paper & \url{https://www.mdpi.com/1099-4300/27/1/62} & 13/01/25\\
    PubMed paper & \url{https://jamanetwork.com/journals/jamanetworkopen/fullarticle/2827327} & 13/01/25\\
    PubMed paper & \url{https://pmc.ncbi.nlm.nih.gov/articles/PMC9568596/pdf/41598_2022_Article_22228.pdf} & 13/01/25\\
    PubMed paper & \url{https://pmc.ncbi.nlm.nih.gov/articles/PMC7037716/pdf/ijerph-17-01062.pdf} & 13/01/25\\
    PubMed paper & \url{https://pmc.ncbi.nlm.nih.gov/articles/PMC5897824/pdf/rsta20160452.pdf} & 13/01/25\\
    PubMed paper & \url{https://journals.physiology.org/doi/epdf/10.1152/japplphysiol.00342.2024} & 21/01/25\\
    PubMed paper & \url{https://pmc.ncbi.nlm.nih.gov/articles/PMC10986173/pdf/fresc-05-1303094.pdf} & 21/01/25\\
    Report & \url{https://creativecommons.org/wp-content/uploads/2024/04/2023-Creative-Commons-Annual-Report-2-1.pdf} & 10/01/25 \\
    Report & \url{https://creativecommons.org/wp-content/uploads/2024/04/240404Towards_a_Books_Data_Commons_for_AI_Training.pdf} & 10/01/25\\
    Report & \url{https://www.lem.sssup.it/WPLem/odos/odos_report_2.pdf} & 10/01/25\\
    User manual & \url{https://manuals.plus/vwar/dt3-mate-sports-smart-watch-manual} & 10/01/25\\
    User manual & \url{https://cms5.revize.com/revize/cityofsedrowoolley/Departments/Solid%20Waste/CompostGuide.pdf} & 09/01/25 \\
    User manual & \url{https://data.europa.eu/sites/default/files/edp_s1_man_portal-version_4.3-user-manual_v1.0.pdf} & 09/01/25\\
    User manual & \url{https://unfccc.int/files/national_reports/non-annex_i_national_communications/non-annex_i_inventory_software/application/pdf/naiis-user-manual.pdf} & 09/01/25\\
    User manual & \url{https://www.researchgate.net/publication/351037551_A_Practical_Guide_to_Building_OWL_Ontologies_Using_Protege_55_and_Plugins} & 10/01/25\\
    Wikipedia & \url{https://en.wikipedia.org/wiki/Logic} & 09/01/25 \\
    Wikipedia & \url{https://en.wikipedia.org/wiki/Hard_problem_of_consciousness} & 09/01/25\\
    Wikipedia & \url{https://en.wikipedia.org/wiki/Artificial_intelligence} & 09/01/25\\
    Wikipedia & \url{https://en.wikipedia.org/wiki/Lyon} & 09/01/25\\
    Wikipedia & \url{https://en.wikipedia.org/wiki/Louis_XIV} & 09/01/25\\
    \hline
    \caption{Description of the 50 newly collected PDFs for dataset creation.}
\end{longtable}

\section{Parsers' features}
\label{sec:parsers-features}
\begin{table*}[h!]
  \centering
  \begin{tabular}{lccccc}
    \hline
    \textbf{Feature}
    & \begin{tabular}{@{}c@{}}\textbf{ChunkNorris}\\\emph{1.0.5}\end{tabular}
    & \begin{tabular}{@{}c@{}}\textbf{Docling}\\\emph{2.15.1}\end{tabular}
    & \begin{tabular}{@{}c@{}}\textbf{Marker}\\\emph{1.2.7}\end{tabular}
    & \begin{tabular}{@{}c@{}}\textbf{Open-Parse}\\\emph{0.7.0}\end{tabular}
    & \begin{tabular}{@{}c@{}}\textbf{PyPDF}\\\emph{5.1.0}\end{tabular}
    \\
    \hline
    \textbf{Text extraction} & x & x & x & x & x \\
    - keeps font styling & x & x & x & x & \\
    - recombine paragraphs & x & x & x & & \\
    \textbf{Tables parsing}  & & & & & \\
    - if as line vectors & x & x & x & x & \\
    - if suggested structure & & x & x & x & \\
    - if as images & & x & x & x & \\
    \textbf{Handles links} & x & & & & \\
    \textbf{Section headers detection} & x & x & x & & \\
    - with hierarchy & x & & x & & \\
    \textbf{Handles equations} &  & x & x & & \\
    \textbf{Removes page headers/footer} & x & x & x & & \\
    \textbf{Built-in chunking method} & x & x & & x & \\
    
    \hline
  \end{tabular}
  \caption{ \label{tab:parsers-features}
    Features of the various parsers used in this work.
  }
\end{table*}

\newpage
\section{Pipeline results}
\label{sec:pipeline-results}
%%%%%%%%%%%% ONE CHUNK %%%%%%%%%%%%%%%%%%%%%%

\begin{table*}[h!]
  \centering
  \begin{tabular}{l l c c c c c c}
    \toprule
    \multirow{2}{*}{\textbf{Parser}} & \multirow{2}{*}{\textbf{Chunker}} & \multicolumn{2}{c}{\textbf{Snowflake 23M}} & \multicolumn{2}{c}{\textbf{Snowflake 109M}} & \multicolumn{2}{c}{\textbf{Snowflake 305M}} \\
    \cmidrule(lr){3-4} \cmidrule(lr){5-6} \cmidrule(lr){7-8}
                                      &                                   & \textbf{R@10} & \textbf{NDCG@10} & \textbf{R@10} & \textbf{NDCG@10} & \textbf{R@10} & \textbf{NDCG@10}\\
    \midrule
    ChunkNorris & Page & 0.45 & 0.28 & 0.26 & 0.18 & \underline{0.83} & 0.62 \\
    ChunkNorris & RC & 0.45 & 0.30 & 0.28 & 0.19 & 0.80 & 0.61\\
    ChunkNorris & ChunkNorris & \textbf{0.57} & \textbf{0.35} & \textbf{0.38} & \textbf{0.21} & \textbf{0.86} & \textbf{0.66} \\
    Docling & Page & 0.45 & 0.30 & 0.24 & 0.16 & 0.81 & 0.62\\
    Docling & RC & 0.36 & 0.25 & 0.22 & 0.14 & 0.76 & 0.57 \\
    Docling & Docling & 0.44 & 0.29 & 0.26 & 0.18 & 0.81 & 0.61\\
    Marker & Page & 0.43 & 0.28 & 0.25 & 0.17 & 0.79 & 0.59\\
    Marker & RC & 0.43 & 0.29 & 0.26 & 0.17 & 0.78 & 0.59 \\
    Open-Parse-P & Page & 0.45 & 0.30 & \underline{0.29} & \underline{0.20} & 0.77 & 0.59\\
    Open-Parse-P & RC & \underline{0.48} & \underline{0.31} & \underline{0.29} & \underline{0.20} & 0.73 & 0.57 \\
    Open-Parse-P & Open-Parse & 0.38 & 0.26 & 0.21 & 0.14 & 0.73 & 0.58 \\
    Open-Parse-U & Page & 0.45 & \underline{0.31} & \underline{0.29} & 0.19 & 0.78 &  0.60\\
    Open-Parse-U & RC & 0.44 & \underline{0.31} & 0.27 & 0.19 & 0.74 & 0.57 \\
    Open-Parse-U & Open-Parse & 0.35 & 0.25 & 0.19 & 0.13 & 0.70 & 0.55 \\
    PyPDF & Page & 0.40 & 0.27  & 0.21 & 0.15 & \underline{0.83} & \underline{0.63} \\
    PyPDF & RC & 0.43 & 0.29 & 0.24 & 0.16 & 0.81 & 0.61 \\
    \bottomrule
  \end{tabular}

  \vspace{1em}

  \begin{tabular}{l l c c c c c c}
    \toprule
    \multirow{2}{*}{\textbf{Parser}} & \multirow{2}{*}{\textbf{Chunker}} & \multicolumn{2}{c}{\textbf{BGE 33M}} & \multicolumn{2}{c}{\textbf{BGE 109M}} & \multicolumn{2}{c}{\textbf{BGE 335M}} \\
    \cmidrule(lr){3-4} \cmidrule(lr){5-6} \cmidrule(lr){7-8}
                                      &                                   & \textbf{R@10} & \textbf{NDCG@10} & \textbf{R@10} & \textbf{NDCG@10} & \textbf{R@10} & \textbf{NDCG@10} \\
    \midrule
    ChunkNorris & Page & 0.75 & 0.57 & 0.73 & 0.56 & 0.79 & 0.60\\
    ChunkNorris & RC & 0.75 & 0.57 & 0.73 & 0.56 & 0.77 & 0.60\\
    ChunkNorris & ChunkNorris & \textbf{0.84} & \textbf{0.65} & \underline{0.81} & \underline{0.61} & \textbf{0.86} & \textbf{0.68} \\
    Docling & Page & 0.76 & 0.57 & 0.75 & 0.57 & 0.77 & 0.59\\
    Docling & RC & 0.75 & 0.56 & 0.73 & 0.56 & 0.76 & 0.60\\
    Docling & Docling & \underline{0.80} & \underline{0.61} & \textbf{0.83} & \textbf{0.63} & \underline{0.82} & \underline{0.65} \\
    Marker & Page & 0.77 & 0.58 & 0.76 & 0.56 & 0.77 & 0.59\\
    Marker & RC & 0.77 & 0.58 & 0.76 & 0.56 & 0.77 & 0.60 \\
    Open-Parse-P & Page & 0.73 & 0.55 & 0.71 & 0.53 & 0.73 & 0.55 \\
    Open-Parse-P & RC & 0.72 & 0.55 & 0.71 & 0.52 & 0.72 & 0.55 \\
    Open-Parse-P & Open-Parse & 0.73 & 0.57 & 0.71 & 0.56 & 0.73 & 0.61\\
    Open-Parse-U & Page & 0.71 & 0.53 & 0.69 & 0.52 & 0.73 & 0.55\\
    Open-Parse-U & RC & 0.71 & 0.54 & 0.69 & 0.52 & 0.73 & 0.56\\
    Open-Parse-U & Open-Parse & 0.69 & 0.53 & 0.68 & 0.53 & 0.71 & 0.58\\
    PyPDF & Page & 0.78 & 0.59 & 0.77 & 0.59 & 0.79 & 0.61 \\
    PyPDF & RC & 0.79 & 0.60 & 0.74 & 0.57 & 0.79 & 0.62 \\
    \bottomrule
  \end{tabular}
  \caption{Comparison of parsers' and chunkers' performance on the IR task depending on various embedding models with the single-chunk dataset. RC stands for Recursive Character chunker and R for Recall.}
  \label{tab:comparison-single}
\end{table*}

%%%%%%%%%%%% MULTI %%%%%%%%%%%%%%%%%%%%%%
\begin{table*}
  \centering
  \begin{tabular}{l l c c c c c c}
    \toprule
    \multirow{2}{*}{\textbf{Parser}} & \multirow{2}{*}{\textbf{Chunker}} & \multicolumn{2}{c}{\textbf{Snowflake 23M}} & \multicolumn{2}{c}{\textbf{Snowflake 109M}} & \multicolumn{2}{c}{\textbf{Snowflake 305M}} \\
    \cmidrule(lr){3-4} \cmidrule(lr){5-6} \cmidrule(lr){7-8}
                                      &                                   & \textbf{R@10} & \textbf{NDCG@10} & \textbf{R@10} & \textbf{NDCG@10} & \textbf{R@10} & \textbf{NDCG@10}\\
    \midrule
    ChunkNorris & Page & \underline{0.42} & 0.27 & 0.32 & \underline{0.22} & 0.72 & 0.50 \\
    ChunkNorris & RC & 0.39 & 0.25 & \underline{0.33} & 0.21  & 0.70 & 0.49 \\
    ChunkNorris & ChunkNorris & \textbf{0.63} & 0.30 & \textbf{0.40} & \underline{0.22} & \underline{0.74} & \textbf{0.56} \\
    Docling & Page & 0.34 & 0.20 & 0.20 & 0.13 & 0.71 & 0.49\\
    Docling & RC & 0.18 & 0.11  & 0.17 & 0.10 & 0.51 & 0.34 \\
    Docling & Docling & 0.35 & 0.20 & 0.15 & 0.12 & 0.59 & 0.44\\
    Marker & Page & 0.31 & 0.20 & 0.22 & 0.14 & 0.62 & 0.43\\
    Marker & RC & 0.32 & 0.21 & 0.25 & 0.15 & 0.61 & 0.44 \\
    Open-Parse-P & Page & 0.39 & \textbf{0.36} & 0.31 & \textbf{0.23} & 0.72 & \underline{0.54}\\
    Open-Parse-P & RC & 0.39 & \underline{0.33} & 0.30 & 0.21 & 0.71 &  0.52 \\
    Open-Parse-P & Open-Parse & 0.28 & 0.20 & 0.18 & 0.12 & 0.39 & 0.31 \\
    Open-Parse-U & Page & 0.40 & 0.30 & 0.30 &  0.21 & \textbf{0.75} & \underline{0.54}\\
    Open-Parse-U & RC & 0.36 & 0.28 & 0.28 & 0.18 & 0.71 & 0.50\\
    Open-Parse-U & Open-Parse & 0.25 & 0.17 & 0.17 &  0.10 & 0.42 & 0.32\\
    PyPDF & Page & 0.24 & 0.18 & 0.15 & 0.13 & 0.70 &  0.51\\
    PyPDF & RC & 0.24 & 0.18 & 0.17 & 0.13 & 0.72 & 0.51 \\
    \bottomrule
  \end{tabular}

  \vspace{1em} 

  \begin{tabular}{l l c c c c c c}
    \toprule
    \multirow{2}{*}{\textbf{Parser}} & \multirow{2}{*}{\textbf{Chunker}} & \multicolumn{2}{c}{\textbf{BGE 33M}} & \multicolumn{2}{c}{\textbf{BGE 109M}} & \multicolumn{2}{c}{\textbf{BGE 335M}} \\
    \cmidrule(lr){3-4} \cmidrule(lr){5-6} \cmidrule(lr){7-8}
                                      &                                   & \textbf{R@10} & \textbf{NDCG@10} & \textbf{R@10} & \textbf{NDCG@10} & \textbf{R@10} & \textbf{NDCG@10} \\
    \midrule
    ChunkNorris & Page & 0.58 & 0.44 & 0.60 & 0.42 & 0.69 & 0.50\\
    ChunkNorris & RC & 0.52 & 0.41 & 0.52 & 0.40 & 0.65 & 0.50\\
    ChunkNorris & ChunkNorris & \textbf{0.78} & \textbf{0.56} & \textbf{0.79} & \textbf{0.52} & \textbf{0.81} & \textbf{0.59} \\
    Docling & Page & 0.56 & 0.41 & 0.54 & 0.40 & 0.68 & 0.48\\
    Docling & RC & 0.48 & 0.35 & 0.56 & 0.39 & 0.57 & 0.42\\
    Docling & Docling & \underline{0.66} & \underline{0.48} & 0.60 & \underline{0.49} & 0.67 & \underline{0.53}\\
    Marker & Page & 0.60 & 0.42 & \underline{0.64} & 0.41 & \underline{0.72} & 0.48\\
    Marker & RC & 0.62 & 0.44 & 0.58 & 0.40 & 0.70 & 0.49\\
    Open-Parse-P & Page & 0.63 & 0.46 & 0.58 & 0.42 & 0.68 & 0.50\\
    Open-Parse-P & RC & 0.62 & 0.46 & 0.55 & 0.43 & 0.71 & 0.51\\
    Open-Parse-P & Open-Parse & 0.44 & 0.34 & 0.43 & 0.34 & 0.41 & 0.34 \\
    Open-Parse-U & Page & 0.62  & 0.44 & 0.58 &  0.43 & 0.68 & 0.50\\
    Open-Parse-U & RC & 0.60 & 0.42 & 0.53 & 0.42 & 0.70 & 0.50\\
    Open-Parse-U & Open-Parse & 0.43 & 0.33 & 0.48 & 0.36 & 0.47 & 0.37 \\
    PyPDF & Page & 0.61 & 0.44 & 0.61 & 0.43 & 0.65 & 0.45\\
    PyPDF & RC & 0.56 & 0.43 & 0.59 & 0.42 & 0.66 & 0.46 \\
    \bottomrule
  \end{tabular}
  \caption{Comparison of parsers' and chunkers' performance on the IR task depending on various embedding models with the multi-chunk dataset. RC stands for Recursive Character chunker and R for Recall.}
  \label{tab:comparison-multi}
\end{table*}

\newpage
\twocolumn
\section{Energy consumption on another hardware}

\label{sec:energy-conso-local}
\begin{table}[h!]
  \centering
  \begin{tabular}{lcc}
    \hline
    \textbf{Parsers} & \begin{tabular}{@{}c@{}}\textbf{CPU}\\ \textbf{energy}\end{tabular} & \begin{tabular}{@{}c@{}}\textbf{GPU}\\ \textbf{ energy}\end{tabular}\\
    \hline
    ChunkNorris  & \underline{0.24Wh} & \textbf{0.0Wh}\\
    Docling & 2.46Wh & \underline{8.61Wh} \\
    Marker & 10.76Wh & 30.10Wh \\
    Open-Parse-P & 1.94Wh & \textbf{0.0Wh}\\
    Open-Parse-U & 44.34Wh & 391.89Wh \\
    PyPDF & \textbf{0.17Wh} & \textbf{0.0Wh}\\
    \hline
  \end{tabular}
  \caption{
    Energy consumption required to parse the 100 PDFs (5286 pages). Hardware: CPU \emph{13th Gen Intel(R) Core(TM) i7-13620H}; GPU \emph{ NVIDIA GeForce RTX 4060 Laptop}; RAM 16 GiB.}
  \label{tab:parsers-conso-total-local-env}
\end{table}

\begin{table}[h!]
  \centering
  \begin{tabular}{lc}
    \hline
    \textbf{Parsers} & \textbf{Parsing time} \\
    \hline
    ChunkNorris & \underline{57ms $\pm$ 165}\\
    Docling & 333ms $\pm$ 519 \\
    Marker & 784ms $\pm$ 637  \\
    Open-Parse-P & 320ms $\pm$ 1080 \\
    Open-Parse-U & 3508ms $\pm$ 5225 \\
    PyPDF & \textbf{46ms $\pm$ 93} \\
    \hline
  \end{tabular}
  \caption{
    Average of the page parsing time for each parser. Hardware: CPU \emph{13th Gen Intel(R) Core(TM) i7-13620H}; GPU \emph{ NVIDIA GeForce RTX 4060 Laptop}; RAM 16 GiB.}
  \label{tab:parsers-time-per-page-local-env}
\end{table}

\end{document}